\documentstyle[12pt]{article}
         \makeatletter
         \def\thefigure{\@arabic\c@figure}\def\fps@figure{tbp}
         \def\ftype@figure{1}\def\ext@figure{lof}
         \def\fnum@figure{\protect\footnotesize Fig.\ \thefigure}
         \def\thetable{\@arabic\c@table}
         \def\fps@table{tbp}\def\ftype@table{2}\def\ext@table{lot}
         \def\fnum@table{\protect\footnotesize Table \thetable}
         \def\@listI{\leftmargin\leftmargini\parsep=0pt\itemsep=0pt}
         \def\thebibliography#1{\section{References}\vspace*{-10pt}\list
          {[\arabic{enumi}]}{\settowidth\labelwidth{[#1]}\leftmargin\labelwidth
          \advance\leftmargin\labelsep
          \usecounter{enumi}}
          \def\newblock{\hskip .11em plus .33em minus .07em}
          \sloppy\clubpenalty4000\widowpenalty4000
          \sfcode`\.=1000\relax}
         
         \def\@nomath#1{\ifmmode \fi}
         \def\mmycite{\@ifnextchar [{\@tempswatrue\@mmycitex}
             {\@tempswafalse\@mmycitex[]}}
         \def\@mmycitex[#1]#2{\if@filesw\immediate%
         \write\@auxout{\string\citation{#2}}\fi
           \def\@citea{}\@mmycite{\@for\@citeb:=#2\do
             {\@citea\def\@citea{,}\@ifundefined
                {b@\@citeb}{{\bf ?}\@warning
                {Citation `\@citeb' on page \thepage \space undefined}}%
         \hbox{\csname b@\@citeb\endcsname}}}{#1}}
         \def\@mmycite#1#2{{{\scriptsize#1}\if@tempswa , #2\fi}}
         \def\mycite#1{$^{\protect\mmycite{#1}}$}
         \def\@cite#1#2{{#1\if@tempswa , #2\fi}}
         \def\thesection {\arabic{section}}
         \def\section#1{\addtocounter{section}{1}\setcounter{subsection}{0}
              \bigskip\medskip{\noindent\bf\thesection.\ #1}
              \medskip}
         \def\thesubsection {\arabic{section}.\arabic{subsection}}
         \def\subsection#1{\addtocounter{subsection}{1}
              \medskip{\noindent\thesubsection.\ #1}
              \medskip}
         \makeatother
\begin{document}
\begin{flushright}
McGill/95-09\\ March 1995
\end{flushright}
\vspace*{-24pt}
\vspace*{0.3in}
\begin{center}
  {\bf Heavy Resonance Production\\
       in Ultrarelativistic Nuclear Collisions$^*$}\\
  \bigskip
  \bigskip
  D. SEIBERT$^{\dag}$\\
  {\em Department of Physics\\
       McGill University, Montreal, QC H3A 2T8, CANADA}
  \bigskip
\end{center}
\smallskip
{\footnotesize
\centerline{ABSTRACT}
\begin{quotation}
\vspace{-0.10in}
Are heavy quarks produced thermally or only by hard parton collisions?
What is the probability that a produced heavy quark or antiquark is
observed in a given resonance?
\end{quotation}}

\section{Introduction}

In this talk, I discuss the problems of heavy quark production and of
heavy resonance production, i.e., are heavy quarks produced thermally or
only by hard parton collisions, and what is the probability that a
produced heavy quark or antiquark is observed in a given resonance?  My
collaborators are Tanguy Altherr at CERN\mycite{as}, who has since
died tragically in a climbing accident, and George Fai at Kent State
University\mycite{sf}.

We use very simple ultrarelativistic nuclear scenarios, basically as
proposed long ago by Bjorken\mycite{bj}.  We assume that the collisions
can be divided into the following periods:
\begin{enumerate}
\item hard parton collisions, in which essentially all of the final-state
entropy is produced, for which the typical time scale is about $0.1$ fm
(we use the standard high energy conventions that $\hbar=c=k_B=1$);
\item thermal equilibration\mycite{hotglue}, which ends after $0.2-0.3$
fm, and possibly even sooner\mycite{gkce};
\item expansion of thermally equilibrated gluon plasma\mycite{hotglue}
(GP), or possibly quark gluon plasma (QGP);
\item phase transition, when the temperature $T \simeq 1$ fm$^{-1}$;
\item resonance gas (RG) expansion; and
\item freezeout, after which the particles leave the hot matter and flow
conveniently to the detectors.
\end{enumerate}
Of course, it is possible that the process of thermal equilibration will
coincide with some of the later equilibrium processes, but we believe
that this will produce only small changes to most of our results.  It is
also possible that there is no phase transition but instead some fast
but smooth change to the entropy and energy densities as a function of
$T$, but again that should not produce large changes in our results.

We make the following approximations to model the dynamics.  First, we
assume that the hot matter is cylindrically symmetric, so
\begin{equation}
f(p_x, p_y, p_z) = f(p_T=\sqrt{p_x^2+p_y^2}, p_z),
\end{equation}
where $f$ is any distribution function.
Second, we assume approximate boost-invariance as observed in proton
collisions\mycite{bj},
\begin{equation}
f(\tau, p_T, y) = f(\tau, p_T).
\end{equation}
Finally, we assume that the hot matter does not expand transversely
during the collision, so that the volume at proper time $\tau$ is
\begin{equation}
\frac {dV(\tau)} {dy} = \pi R^2 \tau,
\end{equation}
where $R$ is the nuclear radius.

We also occasionally assume approximate entropy conservation,
\begin{equation}
\pi R^2 \tau s \simeq 3.6 \, dN/dy, \label{eScon}
\end{equation}
where $s$ is the entropy density and $dN/dy$ is the number of produced
particles per unit rapidity.  In that case, we treat the hot matter as an
ideal gas with zero chemical potential, so
\begin{equation}
s(T) = \sum_i \frac {g_i} {(2\pi)^3T} \int d^3p \frac
{E_i(p) + k^2/3E_i(p)} {e^{E_i(p)/T} \pm 1}. \label{esT}
\end{equation}
Combining eqs.~[\ref{eScon}] and [\ref{esT}], we then estimate $T$ as a
function of $\tau$ for given $dN/dy$.

\section{Heavy quark production}

In this section I review the theory of heavy quark production.  There
have been three types of calculations of heavy quark production in
ultrarelativistic heavy ion collisions.  The initial production cross
sections have been estimated from perturbative quantum chromodynamics
(QCD), either by simply taking cross sections for proton
collisions\mycite{PQCD} and scaling by $A^2$, where $A$ is the number of
nucleons in the nucleus, or in a more sophisticated manner by using the
nuclear (instead of the bare nucleon) parton structure
functions\mycite{sv,lg}.  The results obtained with these two methods do
not differ very much.

Thermal production cross
sections have been calculated by taking the production matrix elements,
convoluting with the four-volume element and the thermal distribution
functions, and integrating over assumed collision
histories\mycite{rm,msm,s,bdmtw,as}.  These thermal calculations have
grown in sophistication with time, but the results have not changed very
much.  Our major addition to these calculations has been the inclusion of
heavy quark production through thermal gluon decay, which has been
previously neglected but is the dominant term in the weak coupling
limit\mycite{as}, due to the anomalously large thermal gluon
width\mycite{Pis}.  For typical values of the strong coupling constant
probed in nuclear collisions, this new production term is comparable to
those previously calculated, although it does not dominate\mycite{bcdh}.

Finally, there is the parton cascade model\mycite{gkce,pcms,gscb}, which
attempts to include both of the previous calculations by using a
perturbative QCD cascade, in which interactions are more or less arbitrarily
cut off at some lower momentum so that cross sections remain finite.
This model seems to work reasonably well for strange ($s$) quark production,
which at least seems to be correctly predicted for proton collisions and
scales with projectile and target in a reasonable manner.  Results for
charm ($c$) and bottom ($b$) production are probably not reliable, since
proton collision results are not reproduced and the cross sections scale
as $A^{5/3}$ in proton-nucleus collisions, instead of $A$ as observed by
experimenters.

In my opinion, the best density estimates and dominant quark production
mechanisms for an Au+Au collision at RHIC energy ($\sqrt{s}=200$
GeV/nucleon) are:
\begin{itemize}
\item[$s$:~] Pure thermal production gives $dN_s/dy \simeq 100$, the
parton cascade model gives $dN_s/dy \simeq 50$, so I expect that
$dN_s/dy \simeq 100$, mostly from thermal collisions.
\item[$c$:~] Thermal production gives $dN_c/dy \simeq 1$, perturbative QCD
gives $dN_c/dy \simeq 1$, so I expect that $dN_c/dy \simeq 1$ with
roughly equal production from hard parton collisions and thermal
collisions.
\item[$b$:~] Thermal production gives $dN_b/dy < 10^{-3}$, perturbative
QCD gives $dN_b/dy \simeq 0.02$, so I expect $dN_b/dy \simeq 0.02$ mostly
from hard parton collisions.
\end{itemize}

\section{Freezeout conditions}

Once we have our heavy quarks, the next question is, ``When do they
freeze out?''  We attempted to solve this problem with a simple model of
heavy quark production and dynamics in an ultrarelativistic nuclear
collision\mycite{sf}.  We first assume that all heavy quark-antiquark
pairs are produced at time $\tau=0$; this is reasonable for $c$ and $b$
quarks even if they are produced thermally, since thermal production is
also concentrated at early times, although it is not such a good
assumption for $s$ quarks.  We then assume that these quarks quickly
thermalize, and that their subsequent trajectory is a random walk in a
thermal bath, with collision frequencies taken from Pisarski\mycite{Pis};
the bath temperature is estimated by assuming entropy conservation.
Finally, we say that the quarks have frozen out when either (i) their
mean free path is larger than the mean distance to leave the hot matter,
or (ii) their random walk has taken them out of the hot matter.

This is of course complicated by the fact that we expect a phase
transition to RG during the evolution of the hot matter.  We take this
into account by assuming that the interaction cross section per unit
entropy is approximately the same in the two phases, and that in the RG
phase the quarks are contained in mesons which we model as
$q\overline{q}$ pairs to estimate the mesonic mean free paths.  We thus
obtain different results by varying the transition temperature, $T_c$,
and the ratio of the number of degrees of freedom in the two phases,
$\nu$.

I show results in Table 1 for S+S collisions at SPS energy ($\sqrt{s}=20$
GeV/nucleon), and for Au+Au collisions at SPS, RHIC, and LHC
($\sqrt{s}=7$ TeV/nucleon) energies, averaged over
$T_c=150$ and $200$ MeV, and over $v=5$ and $10$.  Here $T_f$ is the
freezeout temperature, $\tau_f$ is the proper time, $r_Q$ is the
distance the quark has moved from its point of creation, and $n_f$ is
the mean number of collisions before freezeout.  This last is the most
important number -- when $n_f \simeq 1$ (S+S collisions at SPS),
statistical recombination models should not be expected to work very
well, while when $n_f \gg 1$ (Au+Au collisions at all energies
considered) we expect that statistical models should describe heavy
resonance data very well.

\begin{table}
\begin{center}
\begin{tabular}{lrrccrrrr} \hline \hline
 $A$ & $dN/dy$ & $Q$ &&& $T_f/T_c$ & $\tau_f$ (fm/$c$) & $r_Q$ (fm) & $n_f$ \\
 \hline
  32 &   85    & $s$ &&& 1.0c      & 2.0      & 2.5   &  1    \\
     &         & $c$ &&& 1.0c      & 3.6      & 2.5   &  1    \\
     &         & $b$ &&& 0.8c      & 8.3      & 2.7   &  1    \\[12pt]
 197 & 1000    & $s$ &&& 0.9b      &  15      & 8.8   &  4    \\
     &         & $c$ &&& 0.9b      &  26      & 8.7   &  5    \\
     &         & $b$ &&& 0.7b      &  56      & 8.6   &  5    \\[12pt]
 197 & 2000    & $s$ &&& 1.0w      &  21      & 9.3   &  7    \\
     &         & $c$ &&& 0.9w      &  37      & 9.3   &  8    \\
     &         & $b$ &&& 0.7w      &  78      & 9.3   & 10    \\[12pt]
 197 & 3500    & $s$ &&& 1.0w      &  26      & 9.3   & 10    \\
     &         & $c$ &&& 1.0w      &  45      & 9.3   & 11    \\
     &         & $b$ &&& 0.8w      &  97      & 9.3   & 15    \\
 \hline \hline
 \end{tabular}
 \end{center}
 \caption{Freezeout parameters, adapted from Ref.~2.}
 \end{table}

\section{Statistical recombination}

Our statistical recombination model\mycite{sf} is very similar to those
used by recent authors\mycite{rcss,lthsr,zl}.  The main difference is
that, in addition to ensuring that the quark and antiquark densities,
respectively $\rho_Q$ and $\rho_{\overline{Q}}$, are conserved at
freezeout, we ensure that the local quark-antiquark density,
$\rho^{(2)}_{Q\overline{Q}}(x,x)$, is also conserved.  This is not
trivial, as typically $\rho^{(2)}_{Q\overline{Q}}(x,x) \gg \rho_Q
\rho_{\overline{Q}}$.  We thus include three chemical potentials:
\begin{enumerate}
\item quark, $\mu_Q$;
\item antiquark, $\mu_{\overline{Q}} \neq -\mu_Q$, since the heavy
quarks are not in chemical equilibrium;
\item pair, $\mu_{Q\overline{Q}}$.
\end{enumerate}
The chemical potential for resonance $i$ is
\begin{equation}
\mu_i = \sum_Q \left( k^{(Q)}_i \mu_Q \,
+ \, k^{(\overline{Q})}_i \mu_{\overline{Q}}
+ \, k^{(Q\overline{Q})}_i \mu_{Q\overline{Q}}
\right). \label{emu}
\end{equation}
Here $k^{(Q)}_i$ ($k^{(\overline{Q})}_i$) is the number of quarks
(antiquarks) of flavor $Q$, and $k^{(Q\overline{Q})}_i$ is the number
of pairs (the smaller of $k^{(Q)}_i$ and $k^{(\overline{Q})}_i$).  We
include all confirmed meson and baryon resonances\mycite{PDG} in our
statistical recombination model.

We estimate the two-particle density (and thus $\mu_{Q\overline{Q}}$)
from the freezeout conditions obtained from our simulation.
\begin{equation}
\frac {\rho^{(2)}_{Q\overline{Q}}}{\rho_Q} = \rho_{\overline{Q}} \,
+ \, \frac {3} {4\pi} \left( \frac {3} {5 \, r_Q^2} \right)^{3/2}.
\end{equation}
Results are shown in Table 2, again averaged over $T_c$ and $\nu$.
Although the corrections to $s$ resonance production from the pair
chemical potential are less than order unity, the corrections to $c$ and
$b$ resonance production are huge, giving orders of magnitude changes in
fugacities.  Thus, this pair chemical potential (or its equivalent) must
be included to calculate $c$ and $b$ resonance production.

\begin{table}
\begin{center}
\begin{tabular}{rrrrclcr} \hline \hline
 $A$ & $dN/dy$ & $Q$ & $dN_Q/dy$ && $\rho_Q$ (fm$^{-3}$) &&
  $\rho^{(2)}_{Q\overline{Q}}/\rho_Q\rho_{\overline{Q}}$ \\ \hline
  32 &   85    & $s$ & 5        && 0.06 &&    1 \\
     &         & $c$ & 0.03     && $2 \times 10^{-4}$    &&   40 \\
     &         & $b$ & 0.0003   && $9 \times 10^{-7}$    && 7000 \\[12pt]
 197 & 1000    & $s$ & 25       && 0.01                  &&    1 \\
     &         & $c$ & 0.3      && $8 \times 10^{-5}$    &&    3 \\
     &         & $b$ & 0.003    && $4 \times 10^{-7}$    &&  500 \\[12pt]
 197 & 2000    & $s$ & 50       && 0.016                 &&    1 \\
     &         & $c$ & 1        && $2 \times 10^{-4}$    &&    2 \\
     &         & $b$ & 0.02     && $2 \times 10^{-6}$    &&   80 \\[12pt]
 197 & 3500    & $s$ & 250      && 0.06                  &&    1 \\
     &         & $c$ & 5        && $7 \times 10^{-4}$    &&    1 \\
     &         & $b$ & 0.4      && $3 \times 10^{-5}$    &&    6 \\
 \hline \hline
\end{tabular}
\end{center}
\caption{Freezeout densities, adapted from Ref.~2.}
\end{table}

Finally, this recombination model can be used to gain insight into
predicted phenomena such as suppression of J/$\psi$ resonance production
in ultrarelativistic nuclear collisions\mycite{J/psi}.  Predictions for
the fractions of $c\overline{c}$ and $b\overline{b}$ pairs that freeze
out as $c\overline{c}$ and $b\overline{b}$ mesons are given in Tables 3
and 4.  These quantities serve as surrogates for the J/$\psi$ and
$\Upsilon$ suppression, as these resonances are the most common sources
for observed J/$\psi$ and $\Upsilon$ resonances.  It is worth noting that
the heavy mesons are all predicted to freeze out in the RG in
ultrarelativistic nuclear collisions, so their suppression probably depends
most strongly on (and thus carries the most information about) their
interactions with the RG, and not with the QGP; thus, their dynamics is
only weakly dependent on the presence or absence of QGP.  Otherwise, it is
obvious from the strong dependences on $T_c$ and $\nu$ that the theoretical
uncertainty in these calculations is huge, so much work remains to be done
on the problem of how produced heavy quarks emerge as observed heavy
resonances.

\begin{table}
\begin{center}
\begin{tabular}{rrcllll} \hline \hline
 \multicolumn{2}{l}{$T_c$ (MeV):} &&
 \multicolumn{2}{c}{150} & \multicolumn{2}{c}{200} \\
 $A$ & $dN/dy$ && $\nu$:~5 & 10 & 5 & 10 \\ \hline
  32 &      85 && 0.066 & 0.075  & 0.029 & 0.021  \\
 197 &    1000 && 0.021 & 0.002  & 0.008 & 0.001  \\
 197 &    2000 && 0.009 & 0.004  & 0.004 & 0.001  \\
 197 &    3500 && 0.009 & 0.010  & 0.005 & 0.002  \\ \hline \hline
\end{tabular}
\end{center}
\caption{$c\overline{c}$ meson fractions, from Ref.~2.}
\end{table}

\begin{table}
\begin{center}
\begin{tabular}{rrcllll} \hline \hline
 \multicolumn{2}{l}{$T_c$ (MeV):} &&
 \multicolumn{2}{c}{150} & \multicolumn{2}{c}{200} \\
 $A$ & $dN/dy$ && $\nu$:~5 & 10 & 5 & 10 \\ \hline
  32 &      85 && 0.62  & 0.16   & 0.36  & 0.072  \\
 197 &    1000 && 0.50  & 0.053  & 0.20  & 0.016  \\
 197 &    2000 && 0.20  & 0.013  & 0.070 & 0.005  \\
 197 &    3500 && 0.075 & 0.008  & 0.027 & 0.002  \\ \hline \hline
\end{tabular}
\end{center}
\caption{$b\overline{b}$ meson fractions, from Ref.~2.}
\end{table}

\section{Conclusions}

We have calculated heavy quark production; $s$ quarks are mostly
produced thermally, $b$ quarks are mostly produced by hard parton
collisions, whil $c$ quarks are produced by both mechanisms.  We have
attempted to estimate when freezeout will occur for the various heavy
quark species, and whether there are enough collisions that statistical
recombination models should be reliable guides to heavy resonance
production; the latter  seems to be true for Au+Au collisions at SPS
energy and above, but not necessarily for S+S collisions at SPS.
We found that heavy quarks typically freeze out from the RG, and not
from the QGP, so the relative numbers of the various heavy resonances
are only weakly dependent on the presence or absence of QGP.
Finally, we found that it is necessary to include a pair chemical
potential in statistical recombination models, or to otherwise allow for
the fact that heavy quarks and antiquarks are always produced in
coincidence.

This work was supported in part by the Natural Sciences and Engineering
Research Council of Canada, and in part by the FCAR fund of the
Qu\'ebec government.  It is based on earlier work supported in part by
the U.S. Department of Energy under Grant No.\ DOE/DE-FG02-86ER-40251,
and in part by the North Atlantic Treaty Organization under a Grant
awarded in 1991.

\vspace{1.0in}

\noindent $^*${\em Talk presented at the 11$^{\rm th}$ Winter Workshop
on Nuclear Dynamics, Key West, FL, February 11-18, 1995.}\\
$^{\dag}$Electronic mail: seibert@hep.physics.mcgill.ca.

\end{document}